\documentclass[12pt]{article}
\usepackage{graphicx,amsmath}
\usepackage{units}
\usepackage{color}

\newlength{\dinwidth}
\newlength{\dinmargin}
\def\lapproxeq{\lower .7ex\hbox{$\;\stackrel{\textstyle                                                    
<}{\sim}\;$}}                                                    
\def\gapproxeq{\lower .7ex\hbox{$\;\stackrel{\textstyle                                                    
>}{\sim}\;$}}                                                    
\def\be{\begin{equation}}                                                    
\def\ee{\end{equation}}                                                    
\def\bea{\begin{eqnarray}}                                                    
\def\eea{\end{eqnarray}}

\def\sh{\hat s}
\def\sh2{{\hat s}^2}

\begin{document}

\begin{flushright}                                                    
IPPP/14/55  \\
DCPT/14/110 \\                                                    
\today \\                                                    
\end{flushright} 

\vspace*{0.5cm}

\begin{center}
{\Large \bf The photon PDF of the proton}\\

\vspace*{1cm}
                                                   
A.D. Martin$^a$ and M.G. Ryskin$^{a,b}$  \\                                                    
                                                   
\vspace*{0.5cm}                                                    
$^a$ Institute for Particle Physics Phenomenology, University of Durham, Durham, DH1 3LE \\                                                   
$^b$ Petersburg Nuclear Physics Institute, NRC Kurchatov Institute, Gatchina, St.~Petersburg, 188300, Russia \\          
                                                    
\vspace*{1cm}

\begin{abstract} 
We show how the photon input parton distribution function (PDF) may be calculated with good accuracy, and used in an extended DGLAP global parton analysis in which the photon is treated as an additional point-like parton. The uncertainty of the input photon PDF is relatively small, since the major part of the distribution (which is produced by the coherent emission of the photon from a proton that remains intact) is well known.  We present the expected photon PDFs, and compare the predictions with ZEUS data for isolated photon electroproduction at negative rapidities.

\end{abstract}                                                        
   
 \end{center}

\section{Introduction}
 Precise parton distribution functions (PDFs) are an essential ingredient in analysing data obtained at high energy hadron colliders, such as the Tevatron and the LHC.  In perturbative QCD the PDFs are now known at next-to-next-to leading order (NNLO).  Indeed, with the current precision, it is important to investigate the effect of electroweak corrections in hadron collider physics.
 In particular, the QED contributions have large logarithmic terms, up to $\alpha$log$(Q^2/m_q^2)$, arising from photons emitted from the incoming quark lines.  At high $Q^2$ scales these corrections should be resummed. Fortunately the QCD factorisation theorem also applies to QED, and so the photon-induced logarithms can be absorbed into the PDFs, just as the $\alpha_s$log$Q^2$ terms are summed by DGLAP evolution.  As a consequence the normal DGLAP equations are slightly modified and a photon parton distribution function of the proton, $\gamma^p(x,Q^2)$, is generated. Thus, for example, (at leading order in both $\alpha_s$ and $\alpha$) we have an extra equation for the evolution of the photon PDF
 \be
 \frac{\partial\gamma(x,Q^2)}{\partial{\rm log}Q^2}=\frac{\alpha}{2\pi}\int^1_x\frac{dy}{y}\left(P_{\gamma\gamma}\otimes \gamma+ \sum_1 e_i^2 P_{\gamma q} \otimes q_i \right),
\label{eq:evol}
\ee
where
\be
P_{\gamma\gamma}(y)=-\frac{2}{3}\sum_i e_i^2\delta (1-y),~~~~~~~~
P_{\gamma q}=C_F^{-1} P_{gq}.
\ee
 Similarly, the photon PDF, $\gamma$, contributes to the evolution $\partial q_i /\partial$log$Q^2$ via the $P_{q\gamma}$ splitting.

 \section{Existing determinations of $\gamma^p$ compared to this work}
 
 Indeed, with the present level of precision, it has become topical to treat the photon as one of the point-like partons inside the nucleon and to account for this QED effect explicitly in the global parton analysis. This approach was first followed ten years ago by MRST(2004) \cite{MRST}.  Recently it has been used by the NNPDF \cite{NNPDF} and CTEQ \cite{CTEQ} groups. The central issue is the choice of input distributions for the photon PDFs of the proton and neutron.
 
 In the original MRST study it was assumed that the starting distributions are given by one-photon emission off valence (constituent) quarks in the leading logarithm approximation. For example, for the photon PDF of the proton the starting distribution was taken to be\footnote{Here we write the convolution of quark PDFs and the $P_{gq}$ splitting function explicitly; whereas in \cite{MRST} it was simply denoted by $\otimes$.}
 \be
 \gamma^p(x,Q^2_0)~=~\frac{\alpha}{2\pi}\int\frac{dz}{z}\left[ \frac{4}{9}{\rm log} \left(\frac{Q_0^2}{m^2_u}\right)u_0\left(\frac xz\right)+  \frac{1}{9}{\rm log} \left(\frac{Q_0^2}{m^2_d}\right)d_0\left(\frac xx \right)      \right]
 \frac{1+(1-z)^2}{z},
 \label{eq:2}
 \ee
 where $u_0$ and $d_0$ are the valence-like distributions of the proton, and where the current quark masses were used.  
 
 The most direct measurement of the photon PDF at the time of the MRST analysis appeared to be wide-angle scattering of the photon by a electron beam via the process $ep \to e\gamma X$, where the final state electron and photon are produced with equal and opposite large transverse momentum. The subprocess is QED Compton scattering\footnote{There are other contributions which should be included. These will be discussed in Section \ref{sec:5}.}, $e\gamma \to e\gamma$, for which the contribution to the cross section is
 \be
 \sigma(ep\to e\gamma X)~=~\int dx^\gamma~\gamma^p(x^\gamma,\mu^2)~\hat{\sigma}(e\gamma \to e\gamma),
 \label{eq:comp}
 \ee
 where $\mu$ is the factorization scale.   MRST \cite{MRST} predicted a cross section in agreement with the only measurement of this process available at that time \cite{ZEUS}.
  
 The NNPDF \cite{NNPDF} and CTEQ \cite{CTEQ} groups use a different approach to MRST. They parametrise the input photon PDFs, $\gamma(x,Q_0^2)$, and attempt to determine the parameters from the global data, along with the quark and gluon PDFs.  Unfortunately the present data are not of sufficient accuracy to provide a reasonable determination of the photon input.  
 
 The NNPDF collaboration \cite{NNPDF} used freely parametrised (without bias) starting distributions, including the photon PDFs, and constrain the photon PDFs mainly from the Drell-Yan (low-mass, on-shell $W$ and $Z$ production and high-mass) LHC data. There is expected to be the most sensitivity to the low-mass Drell-Yan data \cite{lowmass}.  However, the uncertainties observed in the resulting photon PDFs are huge, especially at low $x$.  
 
 The preliminary CTEQ analysis \cite{CTEQ} proceeds differently. CTEQ keep a similar theoretical form of the distributions $\gamma(x,Q_0^2)$ to that proposed by MRST, but with an arbitrary normalisation parameter, which is expressed as the momentum fraction, $p_0(\gamma)$, carried by the input photon.
They find that the constraint coming from the energy-momentum sum rule is weak (allowing $p_0(\gamma)$ to range up to 5$\%$), while to fit the updated ZEUS data for $ep \to e\gamma X$ \cite{ZEUS1} requires $p_0(\gamma)\sim 0.1-0.2\%$, using the valence quark induced input (\ref{eq:2}) and allowing for the extra normalisation parameter.
  
\begin{figure} 
\begin{center}
\vspace{-6.0cm}
\hspace{-2.0cm}
\includegraphics[height=13cm]{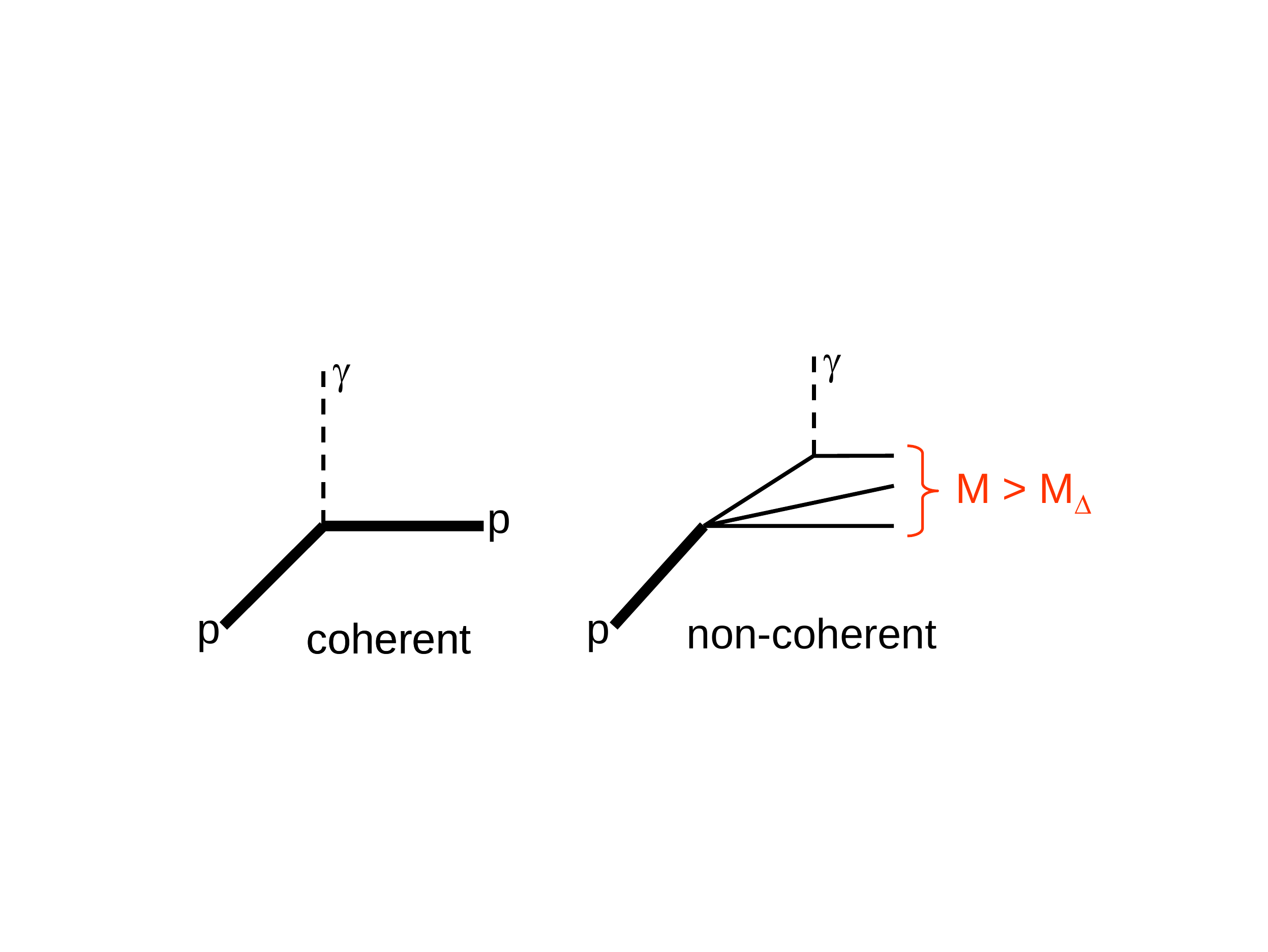}
\vspace{-3.5cm}
\caption{\sf The coherent and incoherent contributions to the photon PDF, $\gamma^p(x,Q^2)$, corresponding, respectively, to photon emission directly from the proton and from a quark.}
\label{fig:1}
\end{center}
\end{figure}
 
 Unlike the above analyses, here we emphasize that the major part of the input, $\gamma^p(x,Q^2_0)$, especially at low $x$, comes from the {\it coherent} emission of the photon from the `elastic' proton, which can be calculated theoretically with good accuracy. The process is sketched in Fig. \ref{fig:1}. (The previous analyses are based only on {\it incoherent} emission from individual quarks within the proton.) The uncertainty on our determination $\gamma^p(x,Q^2_0)= \gamma^p_{\rm coh}+\gamma^p_{\rm incoh}$ comes essentially only from the relatively small  contribution $\gamma^p_{\rm incoh}$ which, from a hadron viewpoint, actually corresponds to the QED excitations of the proton into higher mass states. However, here, in Section \ref{sec:3} we adopt the quark viewpoint, and calculate this contribution as the incoherent emission of photons from quarks within the proton. This contribution turns out to be relatively small. Therefore, since $\gamma^p_{\rm coh}$ may be calculated with good accuracy, it means that the uncertainty in the theoretically determined photon input distributions is expected to be  small; essentially coming from the uncertainty in the `extrapolation' of valence quarks needed to estimate the contribution to $\gamma^p_{\rm incoh}$ coming from the region below the starting scale $Q_0$. We will quantify this uncertainty below.  

 We summarize the discussion of this Section in Table \ref{tab:1}.
 \begin{table} [h]
\begin{center}
\begin{tabular}{|l|l|l|}\hline
 Group &   input photon PDF & data    \\ \hline
  MRST \cite{MRST} & model for $\gamma^p_{\rm incoh}$ & predict $ep \to e\gamma X$   \\
 NNPDF \cite{NNPDF} & freely parametrised & fit to LHC Drell-Yan  \\
 CTEQ \cite{CTEQ} prelim. & parametrise with $p_0(\gamma)$  & fit to $ep \to e\gamma X$ \\
 this work & calculate $\gamma^p_{\rm coh}$ (dominates)  &  predict $ep \to e\gamma X$ \\
    & ~~~~~~~~~ +~model for $\gamma^p_{\rm incoh}$  &   \\
 \hline

\end{tabular}
\end{center}
\vspace*{-0.5cm}
\caption{\sf An outline of the procedure used by the various groups to determine the photon PDF.}
\label{tab:1}
\end{table}

\section{Improved input distributions for the photon PDFs \label{sec:3}}

Here we will follow the MRST approach, but will use much improved starting distributions for the photon PDFs of the proton and neutron. Indeed, we have
\be
\gamma^N(x,Q^2_0)=\gamma^N_{\rm coh}+\gamma^N_{\rm incoh}
\ee
where $N=p,n$. As discussed above, and sketched in Fig. \ref{fig:1}, the contribution $\gamma^p_{\rm coh}$ is caused by coherent photon emission from the proton that remains intact, whereas $\gamma_{\rm incoh}$ is due to non-coherent emission from individual quarks.  
The {\it coherent} emission from the proton is given by \cite{Bud}
\be
\gamma^p_{\rm coh}(x,Q^2_0)=\frac{\alpha^{\rm QED}}{2\pi}~\frac{[1+(1-x)^2]}{x}\int_0^{|t|<Q^2_0}dq^2_t\frac{q^2_t}{(q^2_t+x^2m^2_p)^2}F^2_1(t)\ ,
\label{eq:1}
\ee
where $q_t$ is the transverse momentum of the emitted photon and
\be 
t=-\frac{q^2_t+x^2m^2_p}{1-x}.
\ee
 $F_1$ is the electromagnetic proton form factor corresponding to $\gamma_\mu$ at the vertex.  For the neutron we have
\be
\gamma^n_{\rm coh}(x,Q_0^2)=0.
\ee

For the {\it non-coherent} emission from individual quark lines we use an improved form of (\ref{eq:2}) 
\be
\gamma^p_{\rm incoh}(x,Q^2_0)=\frac{\alpha}{2\pi}\int_x^1 \frac{dz}{z}\left[\frac 49 u_0(\frac xz)+
\frac 19d_0(\frac xz)\right]~\frac{1+(1-z)^2}z\int^{Q^2_0}_{|t_{\rm min}|}~\frac{dt}{t-m_q^2} ~\left( 1-F^2_1(t)\right)\ ,
\label{eq:av}
\ee
where 
\be
t_{\rm min}=-\frac{x}{(1-x)}\left(m^2_\Delta-(1-x)m^2_N\right)
\label{eq:tmin}
\ee
accounts for the fact that the lowest possible proton excitation is the $\Delta$-isobar. The final factor $(1-F^2_1)$ in (\ref{eq:av}) is the probability to have no intact proton in the final state.  We have to exclude an intact proton as its contribution is calculated separately in (\ref{eq:1}).
 
 In (\ref{eq:av}), $m_q=m_d$ when convoluted with $d_0$, and $m_q=m_u$ when convoluted with $u_0$~\footnote{To be precise, we replace the integral $\int dt/(t-m_q^2)$ by $\int \left[\frac{q^2_t/(1-z)}{t-m_q^2}\right]^2\frac{dq^2_t}{q^2_t}$, where $t=t_{\rm min}-q^2_t/(1-z)$ with $t_{\rm min}$ given by (\ref{eq:tmin}).}. In this contribution we use the {\it current} quark masses.
Here the quark distribution $u_0=u_{\rm valence}+2u_{\rm sea}$ is frozen for $Q<Q_0$ at its value at $Q_0$. The same is true for the other quarks - $d,s$.  A similar expression holds for $\gamma^n_{\rm incoh}$, with $4/9 \leftrightarrow 1/9$ and $F_1^p\to F_1^n$.
In this way we get an {\bf upper} limit for the non-coherent contribution to the photon input distributions.
The other extreme is to take for $u_0$ and $d_0$ just the non-relativistic quark model expectation with 
\be
u_0(x)=u_{\rm non-rel}=2\delta(x-1/3)~~~~ {\rm and}~~~~ d_0(x)=d_{\rm non-rel}=\delta(x-1/3) 
\ee
for the proton, and to use {\em constituent} quark masses $m_q=300- 350$ MeV.

The optimum estimate of the non-coherent contribution to the photon PDF input is probably to take a, physics-motivated, linear interpolation between the two limits. That is, to use in (\ref{eq:av}) 
\be
q_0(x,|t|)~=~\frac{|t|}{Q^2_0}~q(x,Q^2_0)+\frac{Q^2_0-|t|}{Q^2_0}~q_{\rm non-rel}(x)
\label{eq:interpol}
\ee
with $m_q=m_{\rm current}+m_{\rm eff}(t)$, where the `effective' constituent quark mass is parametrized by a simplified formula of the form
\be
m_{\rm eff}\simeq m(0)\exp(-b\sqrt{|t|})\ ,
\ee
with $m(0)=345$ MeV and slope $b=1.4$ GeV$^{-1}$ (see, for example,  Fig. 4
in ~\cite{DP}, where the light quark in the instanton vacuum was studied).

In general, one may also account for the $\Delta$-isobar excitation.
In the latter case, we have to add to (\ref{eq:1}) $\gamma^\Delta_{\rm coh}$, which is also of the form of (\ref{eq:1}), but with\footnote{ Here the $F^\Delta(t)$ form factor includes the normalization for $\gamma+p\to \Delta$ cross section, and at small $q_t$ this $p \to \Delta$ transition `form factor' $F^\Delta(t)\propto q_t$ vanishes.} 
\be
F_1(t)/(q^2_t+x^2m^2_p)~~~~{\rm replaced~ by}~~~~ F^\Delta(t)/(q^2_t+x(m^2_\Delta-(1-x)m^2_p).
\ee
For the $\Delta$ contribution 
\be
\label{eq:4}
|t|=\frac{q^2_t+x(m^2_\Delta-(1-x)m^2_N)}{1-x}.
\ee
Also when including the $\Delta$ contribution we have to replace in (\ref{eq:av}) 
\be
[1-F^2_1(t)] ~~~~~{\rm by}~~~~~ [1-F^2_1(t)]-F^2_\Delta(t)\Theta(|t|(1-x)-x(m^2_\Delta-(1-x)m^2_N),
\ee
 where here $|t|$ is given by (\ref{eq:4}).  In addition, it is possible to include a coherent contribution caused by the anomalous magnetic moment of the proton, described by the proton form factor $F_2$.
 These non-logarithmic corrections will reduce the remaining incoherent contribution and therefore decrease the final uncertainty in the input $\gamma^p(x,Q_0^2)$. However, since they do not change the result noticeably, we do not consider these possibilities here.

\section{Results for the photon PDF}
\begin{figure} 
\begin{center}
\vspace{-8.0cm}
\includegraphics[height=16cm]{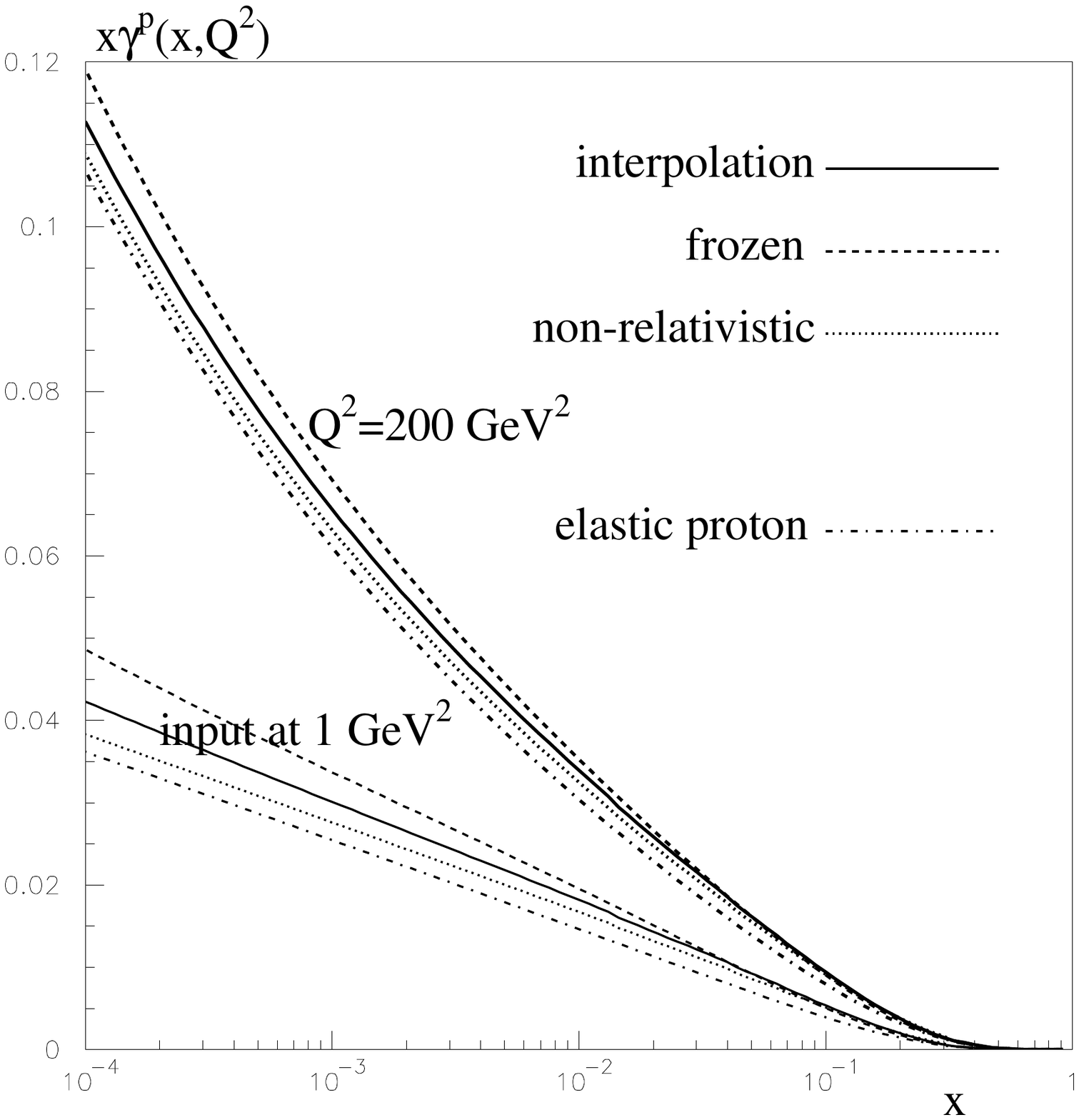}
\vspace{-0.5cm}
\caption{\sf The photon PDF of the proton at input ($Q_0^2=1$ GeV$^2$) and after evolution up to $Q^2=200$ GeV$^2$. At each $Q^2$ value, the lowest curve is $\gamma^p_{\rm coh}$ and the continuous `interpolating' curve (obtained from (\ref{eq:interpol})) is the effect of adding the  $\gamma^p_{\rm incoh}$ contribution. At input, the proton momentum fraction carried by the photon is 0.16$\%$.}
\label{fig:2}
\end{center}
\end{figure}
\begin{figure} 
\begin{center}
\vspace{-6.7cm}
\includegraphics[height=16cm]{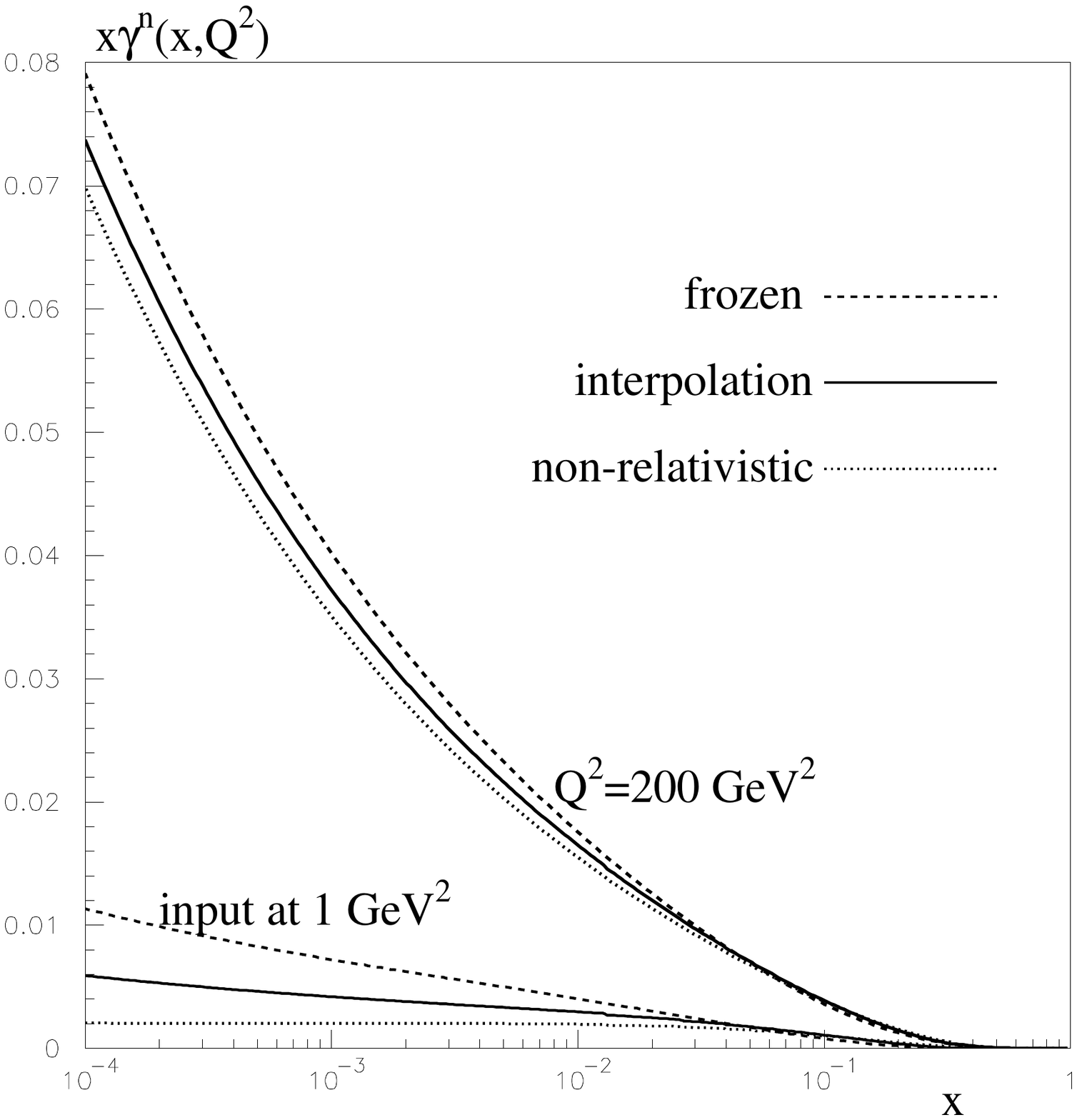}
\vspace{-0.5cm}
\caption{\sf The photon PDF of the neutron at input ($Q_0^2=1$ GeV$^2$) and after evolution up to $Q^2=200$ GeV$^2$. At each $Q^2$ value, the three curves correspond to the upper and lower estimates of $\gamma^n_{\rm incoh}$, together with continuous (`interpolating') curve obtained from (\ref{eq:interpol}).}
\label{fig:2n}
\end{center}
\end{figure} 
\begin{figure} 
\begin{center}
\vspace{-7.0cm}
\includegraphics[height=16cm]{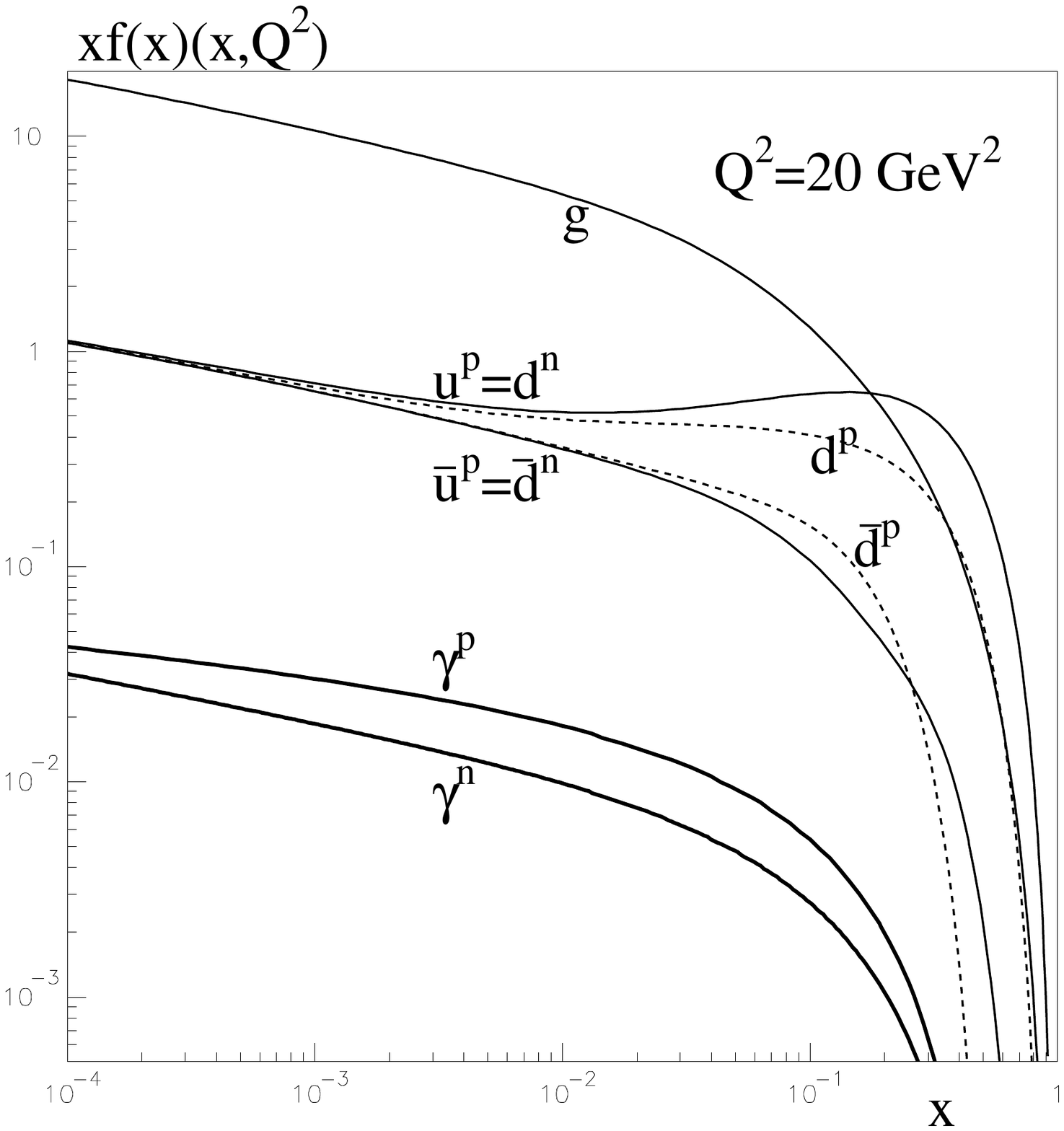}
\vspace{-0.5cm}
\caption{\sf The photon PDFs of the proton and neutron compared with the MSTW \cite{MSTW} NLO quark and gluon distributions at $Q^2=20$ GeV$^2$.}
\label{fig:3n}
\end{center}
\end{figure} 

In Figs. \ref{fig:2} and \ref{fig:2n} we show, respectively, the input distributions of the photon PDF for the proton and neutron at $Q_0^2=1$ GeV$^2$, as calculated in the previous Section, together with the photon PDF determined from
 the evolution up to $Q^2=200$ GeV$^2$ using (\ref{eq:evol}), but with NLO partons.
We see that the major part of the input photon PDF of the proton is generated by  coherent emission of the photon from an `elastic' proton, while the non-coherent contribution only enlarges this value by about 20\%.   Actually, the figure shows three curves for the inclusion of $\gamma^p_{\rm incoh}$, corresponding to the two limits of the input distribution together with their `interpolated average', shown by the continuous curve. 
We also note that for a low $x\sim 10^{-2}\ -\ 10^{-3}$ the evolution to $Q^2=200$ GeV$^2$ increases the photon density by about a factor of two; whereas for $x=10^{-4}$ the increase is about a factor of three.
Considering the two terms involving $P_{\gamma\gamma}$ and $P_{\gamma q}$ of (\ref{eq:evol}), which contribute to the evolution of the photon PDF, we note that the first term, $P_{\gamma\gamma}$, decreases $\gamma^p$ only slightly (less than 1$\%$ in the evolution up to $Q^2=200$ GeV$^2$), whereas the growth comes from the $P_{\gamma q}$ term -- the photons emitted by quarks. We have the same growth of each curve due to the linear nature of DGLAP evolution.
 
  Here we evolve using MSTW\cite{MSTW} NLO quarks\footnote{We find that the use of the updated CPdeut parton set of MMSTWW \cite{MMSTWW} makes a negligible difference.}. At first sight we might expect the contribution generated by gluons, via the gluon-photon splitting, to be important, due to the large gluon PDF, especially at low $x$. However the value of the NLO splitting function $P^{(1)}_{\gamma g}(z)$ is rather small\footnote{We extract the splitting function as the term proportional to $C_F N_f$ from the known $P^{(1)}_{gg} (z)$ splitting \cite{FP}.}.  Moreover, $P^{(1)}_{\gamma g}(z)$ is negative at large $z$. As a result, the inclusion of gluons enlarges $\gamma^p(x,Q^2)$ by less than 2$\%$, and we neglect this effect.

Recall that for the photon PDF of the neutron we have $\gamma^n_{\rm coh}(x,Q^2_0)=0$, and so the input is given entirely by $\gamma^n_{\rm incoh}(x,Q^2_0)$,  
 see Fig. \ref{fig:2n}. However, the {\it increase} in $\gamma^n$ in the evolution up to $Q^2=200$ GeV$^2$ (which is driven by the final term in (\ref{eq:evol})) is comparable to that for $\gamma^p$. In Fig. \ref{fig:3n} we compare $\gamma^p$ and $\gamma^n$ with the other PDFs at $Q^2=20$ GeV$^2$.

\begin{figure} 
\begin{center}
\vspace{-6.0cm}
\vspace{-5.0cm}
\includegraphics[height=24cm]{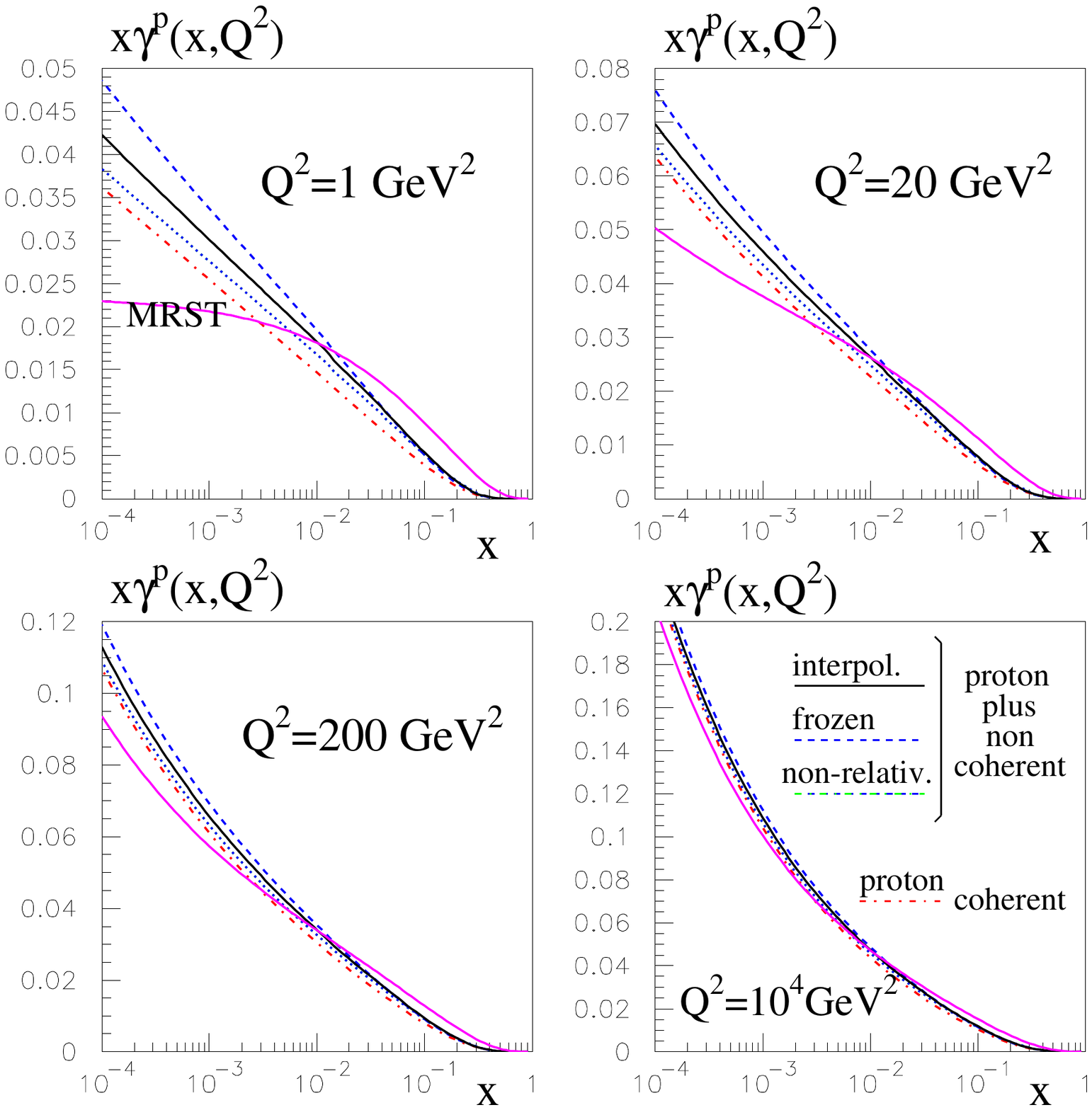}
\vspace{-1.5cm}
\caption{\sf The photon PDF at input $Q_0^2=1$ GeV$^2$ and after evolution up to $Q^2=20,~200, $ and $10^4$ GeV$^2$. The notation of the curves are as in Fig. \ref{fig:2}.  We have included the prediction of MRST(2004) \cite{MRST} for comparison.}
\label{fig:3}
\end{center}
\end{figure} 

In Fig. \ref{fig:3} we show how $\gamma^p$ evolves as $Q^2$ increases from input $Q_0^2=1$ to $Q^2=20,~200$ and $10^4$ GeV$^2$, together with the behaviour predicted\footnote{Actually we use the MRST formulation, but with NLO MSTW partons \cite{MSTW}, which make little change to the behaviour of $\gamma^p$.} by MRST(2004) \cite{MRST} input.  

The comparison of the predictions by MRST \cite{MRST} and of this work, may, at first sight, appear surprising.  MRST is purely based on the incoherent contribution, $\gamma^p_{\rm incoh}$, whereas here the prediction comes dominantly from the coherent contribution (dashed-dotted curve) with a small addition from $\gamma^p_{\rm incoh}$.  The explanation is as follows. The incoherent contribution determined by MRST should be suppressed by $t_{\rm min}$, (\ref{eq:tmin}), and by $[1-F_1^2(t)]$ of (\ref{eq:av}).
In the present work, the coherent emission from the proton is added. The above two effects (that is the suppression of the incoherent contribution and the inclusion of the coherent emission) partly compensate each other.  However, indeed at large $x$, where $|t_{\rm min}|$ is large, MRST goes above the present input, while at low $x$, where
$|t_{\rm min}| \simeq (xm_N)^2$ is even less than the current quark mass, our input exceeds the MRST curve.


\section{Comparison with $ep \to e \gamma X$ data   \label{sec:5}}
\begin{figure} 
\begin{center}
\vspace{-3.0cm}
\vspace{-.0cm}
\includegraphics[height=8cm]{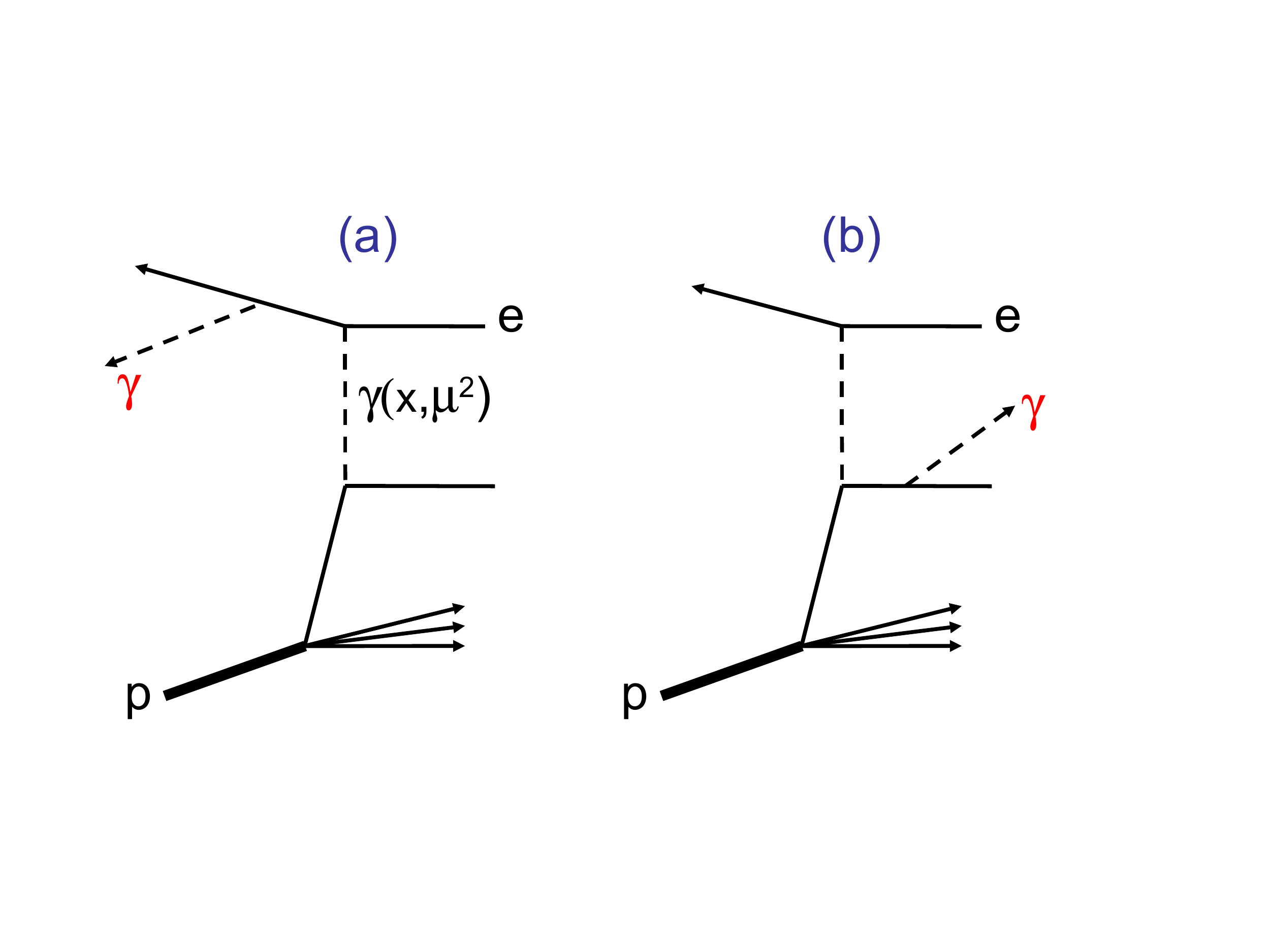}
\vspace{-1.5cm}
\caption{\sf Representative diagrams mediating inclusive  electroproduction of isolated photons, $ep\to e\gamma X$: (a) photon emitted by electron, (b) photon emitted by a quark.}
\label{fig:LLQQ}
\end{center}
\end{figure} 
To probe the photon PDF experimentally we should consider a hard subprocess where the photon distribution dominates. A good example is the inclusive electroproduction of isolated photons observed in the direction of the incoming electron. This reaction is mediated by the $e\gamma \to e\gamma$ hard subprocess
and its cross section is described by (\ref{eq:comp}), and sketched in Fig. \ref{fig:LLQQ}(a). It is known that this Compton scattering process is sharply peaked in the backward direction. Therefore the outgoing photon should be observed at high negative rapidity, $\eta^\gamma$, at angles close to the electron beam. Indeed it was already shown \cite{G1,G2} that this component (called LL) already dominates  for\footnote{Here we adopt the convention for $\eta^\gamma$ used by the ZEUS collaboration \cite{ZEUS1}.} $\eta^\gamma <-1$.

Unfortunately, the present experimental data are quite limited and the lowest rapidity bin is $-0.7<\eta^\gamma<-0.3$.  We compare our theoretical prediction for the LL component with the data in Tables \ref{tab:2} and \ref{tab:3}.  As can be seen from the comparison in the first bin, the predicted cross section is close to the measured value. The result depends on the choice of factorisation scale. We present results for $\mu=E_T^\gamma$ and $E_T^\gamma /2$ to indicate the scale dependence.
\begin{table} [h]
\begin{center}
\begin{tabular}{|c|r|c|c|}\hline
$\eta^\gamma$ range & \multicolumn{3}{c|}  {$d\sigma(ep\to e\gamma X) /d\eta^{\gamma}$  ~~(pb)}  \\ \hline    
 & experiment  &  $\mu=E_T^\gamma$   & $\mu=E_T^{\gamma}/2$ \\
  \hline
 \bf $-0.7--0.3$ &  $17.4~ \pm 0.9 ^{~+0.5}_{~-0.7}$ &16.4 &13.3\\
 $-0.3-0.1$ & $13.0 ~\pm 0.8 ^{~+0.6}_{~-0.3}$  & 7.7&6.3\\
 $0.1-0.5$ & $10.7 ~\pm 0.9 ^{~+0.7}_{~-0.4}$  & 2.7 & 2.24\\
 $0.5-0.9$  &  $8.7 ~\pm 0.9 ^{~+1.1}_{~-0.7}$ & 0.8 &0.65 \\
 \hline

\end{tabular}
\end{center}
\vspace*{-0.5cm}
\caption{\sf The second column gives the values of the $ep\to e\gamma X$ cross section measured by the ZEUS collaboration \cite{ZEUS1} in different rapidity, $\eta^\gamma$, intervals. The final two columns show the contribution to the cross section arising from the LL process of Fig. \ref{fig:LLQQ}(a) for two different choices of the factorisation scale $\mu$.}
\label{tab:2}
\end{table}

 \begin{table} [h]
\begin{center}
\begin{tabular}{|l|l|c|c|}\hline
$E_T^\gamma$ range & \multicolumn{3}{c|}  {$d\sigma(ep\to e\gamma X) /dE_T^{\gamma}$  ~~(pb/GeV)}  \\ \hline 
 (GeV) & experiment  &  $\mu=E_T^\gamma$   & $\mu=E_T^{\gamma}/2$ \\
  \hline
 $4-6$ &  $4.87~ \pm 0.28 ^{~+0.40}_{~-0.23}$ &2.4 &1.95\\
 $6-8$ & $2.40 ~\pm 0.16 ^{~+0.09}_{~-0.11}$  & 1.46&1.22\\
 $8-10$ & $1.24 ~\pm 0.11 ^{~+0.03}_{~-0.04}$  & 0.88 & 0.74\\
 $10-15$  &  $0.55 ~\pm 0.04 ^{~+0.03}_{~-0.03}$ & 0.12  & 0.10 \\
    \hline
\end{tabular}
\end{center}
\vspace*{-0.5cm}
\caption{\sf The second column gives the values of the $ep\to e\gamma X$ cross section measured by the ZEUS collaboration \cite{ZEUS1} in different $E_T^\gamma$ intervals. The final two columns show the contribution to the cross section arising from the LL process of Fig. \ref{fig:LLQQ}(a) for two different choices of the factorisation scale $\mu$.}
\label{tab:3}
\end{table}
  
  At large $\eta^\gamma$, the contribution of the Compton-induced process decreases rapidly. In this domain, inclusive isolated photons are mainly produced by quarks, see Fig. \ref{fig:LLQQ}(b).
  
Note that in our theoretical calculation of $ep\to e\gamma X$ we have accounted for the angular, the $E_T$ and the other experimental cuts imposed by the ZEUS collaboration \cite{ZEUS1}, but we have no possibility to include the photon isolation criteria. Therefore the observed cross section corresponding to the LL process should be lower than our prediction.

In Table \ref{tab:3} we compare our prediction of the LL  contribution with $E_T$ dependence of the measured cross section. However, now the data were collected over a large rapidity interval: $-0.7<\eta^\gamma <0.9$.  Here the quark contribution is important, and the LL subprocess describes only about half of the cross section.

\section{Conclusions}
We have demonstrated that the major part of the photon input PDF of the proton (caused by the coherent emission of the photon that does not destroy the proton) can be calculated with good accuracy. This strongly reduces the possible uncertainties in the QED part of an extended global parton analysis which includes the photon as a point-like parton.  In this way, we evaluate the expected photon PDFs by DGLAP evolution with LO QED splittings and NLO MSTW quarks.  Note that the further step of including the photon-to-quark splitting will introduce a small violation of isospin symmetry in the `singlet' PDF, in particular $u^d \ne d^n$.  The resulting photon distributions agree with data for the electroproduction of isolated photons, $ep\to e\gamma X$, at negative rapidities where the cross section is dominated by the $e\gamma \to e\gamma$ hard subprocess.

\section*{Acknowledgements}

MGR thanks the IPPP at the University of Durham for hospitality. This work was supported by the Federal Program of the Russian State RSGSS-4801.2012.2.  We thank Robert Thorne for valuable discussions.


\end{document}